\documentclass[pre,preprint,amsfonts]{revtex4-1}
\usepackage{graphicx}
\usepackage{tikz,xcolor,hyperref}
\usepackage{epstopdf}
\usepackage{bm}
\usepackage{amsmath,amssymb,amsfonts}
\usepackage{mathtools}
\usepackage[caption=false]{subfig}

\definecolor{mypink}{RGB}{255,0,127}
\definecolor{myblue}{RGB}{0,0,255}
\definecolor{mygreen}{RGB}{102,204,0}
\definecolor{myorange}{RGB}{255,128,0}
\definecolor{mypurple}{RGB}{127,0,255}

\begin{document}
\title{Effect of overlap on spreading dynamics on multiplex networks}

\author{Huan Wang$^{1}$}

\author{Chuang Ma$^{1}$}

\author{Hanshuang Chen$^{2}$}\email{chenhshf@ahu.edu.cn}

\author{Haifeng Zhang$^{1}$}\email{haifengzhang1978@gmail.com}

\affiliation{$^{1}$School of Mathematical Science, Anhui University,
Hefei, 230601, China \\ $^{2}$School of Physics and Materials
Science, Anhui University, Hefei, 230601, China}


\begin{abstract}
In spite of the study of epidemic dynamics on single-layer networks
has received considerable attention, the epidemic dynamics on
multiplex networks is still limited and is facing many challenges.
In this work, we consider the susceptible-infected-susceptible-type
(SIS) epidemic model on multiplex networks and investigate the
effect of overlap among layers on the spreading dynamics. To do so,
we assume that the prerequisite of one $S$-node to be infected is
that there is at least one infectious neighbor in each layer. A
remarkable result is that the overlap can alter the nature of the
phase transition for the onset of epidemic outbreak. Specifically
speaking, the system undergoes a usual continuous phase transition
when two layers are completely overlapped. Otherwise, a
discontinuous phase transition is observed, accompanied by the
occurrence of a bistable region in which a disease-free phase and an
endemic phase are coexisting. As the degree of the overlap
decreases, the bistable region is enlarged. The results are
validated by both simulation and mean-field theory.
\end{abstract}
 \maketitle

\section{Introduction}
In the past decades, complex networks have proved to be a powerful
framework to characterize the interaction among the constituents of
a variety of complex systems, examples range from the social to
technological, biological, and other systems in real world
\cite{newman2018networks}. Up to now, there are a large number of
works paid their attention to the study of the structures of complex
networks and the dynamical behaviors taking place on them
\cite{SIR03000167,PRP06000175,PRP08000093,RevModPhys.87.925,PR.687.1,RMP08001275}.
However, most of these existing achievements mainly focus on
single-layer networks. In fact, many real-world complex systems are
usually composed of multilayer networks \cite{JCN2014,Bianconi2018}.
Multilayer network is a general concept, which includes
interdependent networks, interconnected networks, multiplex
networks, network of network, and so forth. For example, an
interdependent network is formed by the power and communication
infrastructures, and the transportation system including a set of
locations which is connected by roads, railways, waterways, or
airline connections. It has been recognized that the multilayer
networks can present some novel features different from the
single-layer networks, such as complexity, diversity and fragility
\cite{PRP2014,Nature2010,PhysRevLett.109.248701,PhysRevLett.110.028701,NatPhys8.717.2018}.
The researches on multilayer networks have covered a variety of
dynamics including evolutionary games
\cite{Wang_EPJB88.2015.124,NJP19.073017.2017,NJP20.075005.2018},
synchronization
\cite{PhysRevLett.112.248701,SR6.39033.2016,SA2.e1601679.2016},
opinion formation \cite{PhysRevE.89.062818,NJP18.023010.2016},
transportation \cite{PhysRevLett.116.108701,PhysRevLett.120.068301},
and super-diffusive behavior
\cite{PhysRevLett.110.028701,PhysRevE.88.032807,cencetti2019diffusive}.

Epidemic spreading, such as susceptible-infected-susceptible (SIS)
model, is not only a paradigm for studying non-equilibrium phase
transitions, but also has wide applications in real epidemics,
computer viruses, rumor spreading, or signal propagation in neural
networks \cite{hinrichsen2000non}. Therefore, the study of epidemic
spreading on networks is always one of the most active areas in
network science \cite{vespignani2012modelling}. Recently, with the
study in depth of multilayer networks, epidemic spreading on
multilayer networks has also attracted some attention
\cite{IEEE2015,NatPhy2.901.2016,Arruda2018physrep}. Cozzo \emph{et
al.} \cite{PhysRevE.88.050801} have shown that the epidemic
threshold for the SIS model in a multilayer network is always lower
than that in any isolated network. Using an individual-based
mean-field approach, Wang \emph{et al.} \cite{PhysRevE.88.022801}
further showed that the epidemic threshold can be reduced
dramatically if two nodes corresponding to dominant eigenvector
components of the adjacency matrices of isolated networks are
connected. Similar results were also obtained by a degree-based
mean-field approach \cite{PhysRevE.86.026106}. However, Dickison
\emph{et al.} \cite{PhysRevE.85.066109} unveiled, based on the
percolation theory \cite{J.Stat.Mech.2017.034001}, one important
difference between the susceptible-infected-recovered (SIR) model
and the SIS model when the coupling between layers is weak enough.
Spreading processes in structured metapopulations can be well
characterized within the framework of multilayer networks as well
\cite{PhysRevX.8.031039,PhysRevE.87.032809,PhysRevE.90.032806}. de
Arruda \emph{et al.} \cite{PhysRevX.7.011014} used a tensorial
representation \cite{PhysRevX.3.041022} to derive analytical
expressions for the epidemic threshold of the SIS and SIR model on
multilayer networks. They showed, on the one hand, the existence of
disease localization \cite{PhysRevLett.109.128702} and the emergence
of two or more susceptibility peaks. On the other hand, they found
that, when the layer with the lowest eigenvalue is located at the
center of multiplex networks, it can effectively act as a barrier to
the disease.

Multiplex network is a special type of multilayer network, where the
links at each layer represent a different type of interaction
between the same set of nodes. One typical example of the multiplex
network is social networks, where nodes represent individuals and
the different layers correspond to different types of relationship,
such as family, friendships, work-related. Multiplex network also
provides a convenient framework for studying the interplay between
different dynamical processes
\cite{PhysRevLett.118.138302,PhysRevLett.114.038701}, including the
competing spreading process of epidemic and awareness
\cite{PhysRevLett.111.128701,Kan2017}, the cooperative effect among
different spreading dynamics \cite{SR4.5097.2014}, and the interplay
of spreading dynamics and stochastic migration among different
layers \cite{IEEE2017,an2018spontaneous}.

Very recently, discontinuous phase transition of the spreading model
on multiplex networks has received growing attention.
Vel\'asquez-Rojas and Vazquez \cite{PhysRevE.95.052315} coupled
contact process for disease spreading with the voter model for
opinion formation take place on two layers of networks, and they
showed that a continuous transition in the contact process becomes
discontinuous as the infection probability increases beyond a
threshold. Pires \emph{et al.} \cite{J.Stat.Mech.2018.053407}
proposed an SIS-like model with an extra vaccinated state, in which
individuals vaccinate with a probability proportional to their
opinions. Meanwhile, individuals update their opinions in terms of
peer influence. They also observed a first-order active-absorbing
phase transition in the model. Jiang and Zhou
\cite{Sci.Rep.2018.8.1629} studied the effect of resource amount on
epidemic control in a modified SIS model on a two-layer network, and
they found that the spreading process goes through a first-order
phase transition if the infection strength between layers is weak.
Su \emph{et al.} \cite{NJP2018.20.053053} proposed a reversible
social contagion model of community networks that includes the
factor of social reinforcement. They showed that the model exhibits
a first-order phase transition in the spreading dynamics, and that a
hysteresis loop emerges in the system when there is a variety of
initially adopted seeds. Chen \emph{et al.} \cite{NJP20.013007.2018}
studied the dynamics of the SIS model in social-contact multiplex
networks when the recovery of infected nodes depends on resources
from healthy neighbors in the social layer. They found that as the
infection rate increases the infected density varies smoothly from
zero to a finite small value and then suddenly jumps to a high
value, where a hysteresis phenomenon was also observed.

As mentioned in the last paragraph, most of reports on discontinuous
phase transitions in spread models were mainly caused by the
coupling between different dynamics across layers. A natural
question arises: whether such a discontinuous phase transition
appears in a single spreading model on multiplex networks? To the
end, in this work we want to explore a novel discontinuous phase
transition in the SIS model. We propose a variant of the SIS model
on multiplex networks in which a susceptible individual can be
infected only when s(he) has at least one infectious neighbor in
each layer. It is obviously that the model incorporates a
non-additive characteristic of spread dynamics in multiplex
networks. As we shall show later, such a nonlinear effect in
interlayer interactions can induce a discontinuous phase transition
for the onset of epidemic outbreak. It is also known that if the
spreading dynamics is only a simple superposition of those in each
layer, a usual continuous phase transition was observed
\cite{PhysRevE.88.050801,PhysRevX.7.011014}. Moreover, our model is
motivated by some real-world situations. For example, in a rumor
spreading process, a piece of false news is likely to be accepted by
a person if it was shared simultaneously by multiple types of
relationships, such as family members, friends, and coworkers. A
person may prone to purchase a new commodity when (s)he receives
recommendations unanimously from friends of different online
shopping sites \cite{zhang2016dynamics}. The main findings of the
present work is summarized as follows. A key factor to the nature of
phase transition is the degree of edge overlap among different
layers. In particular, when the edges in different layers are
totally overlapping, the model presents a usual continuous phase
transition as the SIS model taking place on the single-layer
networks. Interestingly, when the edges are not totally overlapping,
the model shows a novel discontinuous phase transition, accompanied
by the emergence of bistable region where the endemic extinction
phase and the endemic spread phase are coexisting. The lower degree
of overlap, the wider the bistable region is. We also develop a
mean-field theory to validate the correctness of the results.

\section{Model and Simulation Details}
We consider a spreading model on multiplex networks with two layers,
in which each layer contains the same number $N$ of nodes and there
exists a one-to-one correspondence between nodes in different
layers. The topology in each layer is described by an adjacency
matrix $\textbf{A}^{\ell}$ ($\ell=1,2$), whose entries
$A_{ij}^{\ell}$ are defined as $A_{ij}^{\ell}=1$ if there is an edge
from node $j$ to node $i$ in the $\ell$-th layer, and
$A_{ij}^{\ell}=0$ otherwise. Note that the topology at each layer
may be different. For simplicity, we consider connectivity in each
layer is symmetric, $A_{ij}^{\ell}=A_{ji}^{\ell}$, $\forall i,j$,
and the numbers of edges in two layers are the same,
$M=\sum\nolimits_{i < j} {A_{ij}^1}  = \sum\nolimits_{i < j}
{A_{ij}^2}$. We define the fraction of overlapping edges on two
layers as \cite{PhysRevE.87.062806},
\begin{eqnarray}
\mathcal {O}= \frac{{\sum\nolimits_{i < j} {A_{ij}^1A_{ij}^2}
}}{M},\label{eq1}
\end{eqnarray}
with $0 \leqslant \mathcal {O} \leqslant 1$. For $\mathcal {O}=0$,
there is no overlapping edge in two layers, and \textbf{for
$\mathcal {O}=1$, the topologies in two layers are exactly the
same}. To generate a duplex network with a given $\mathcal {O}$, we
first produce two identical networks as the first layer and the
second layer, respectively. Then, we fix the first layer unchanged
and rewire the edges in the second layer. The rewiring process is
described as follows \cite{PhysRevE.97.042314}. The first step is to
randomly choose an overlapping edge in the two layers. The second
step is to break the edge and then to randomly generate a new edge
in the second layer, in which we ensures that the new edge in the
second layer does not overlap with the first layer. Repeat this
process many times until a given value of $\mathcal {O}$ is reached.

\begin{figure}
\centerline{\includegraphics*[width=0.8\columnwidth]{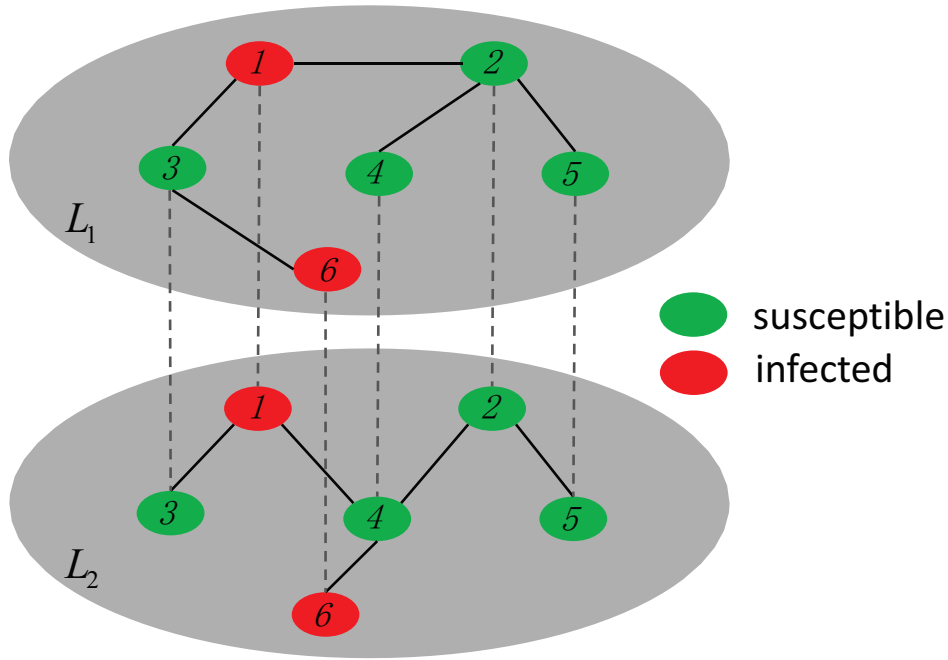}}
\caption{A schematic of our model. A susceptible node can be
infected only if there are at least one infected neighbor in each
layer, and an infected node can spontaneously recover to be
susceptible. According to Eq.(2), node 3 can be infected by the
common infected neighbor (node 1) with the rate $1.5 \lambda$.  But
node 4 cannot be infected as (s)he has no infectious neighbors in
the first layer. \label{fig1}}
\end{figure}

We consider an SIS-type spreading dynamics on duplex networks. Each
node is either susceptible ($\sigma_i(t)=0$) or infected
($\sigma_i(t)=1$) at time $t$. The dynamics of the model is defined
as follows. (i) \emph{Infection}: For a susceptible node $i$, (s)he
can be infected only if there are at least one infectious neighbor
in each layer. Denoting by $n_i^\ell=\sum\nolimits_j {A_{ij}^l}
{\sigma _j}$ the number of infectious neighbors of node $i$ in the
$\ell$-th layer, the rate of node $i$ being infected at time $t$ can
be written as
\begin{eqnarray}
{R_{\inf }} = \lambda \left( {\frac{{n_i^1 + n_i^2}}{2}}
\right)\Theta \left( {n_i^1 - 1} \right)\Theta \left( {n_i^2 - 1}
\right),\label{eq2}
\end{eqnarray}
where $\lambda$ is the infection rate, and $\Theta \left( x \right)$
is the Heaviside function defined as $\Theta \left( x \right)=1$ for
$x\geq0$ and $\Theta \left( x \right)=0$ for $x<0$. The Heaviside
function in Eq.(\ref{eq2}) renders that the total spreading rate is
not a simple superposition of the spreading rates in two layers. As
mentioned before, we have shown that the setting of Eq.(\ref{eq2})
incorporates some practical considerations observed in real
situations, such as rumor spread and commodity recommendations,
which highlights the importance of social reinforcement in the
spreading of information \cite{centola2010spread}. We should also
note that the spreading dynamics in our model is similar to the
threshold model \cite{watts2002simple} and core spreading model
\cite{NJP17.023039,PhysRevE.87.062819} in single-layer networks.
(ii) \emph{Recovery}: For an infectious node $i$, (s)he becomes
spontaneously susceptible at time $t$ with a recovery rate $\mu$.
Without loss of generality, we set to $\mu=1$ and define
$\beta=\lambda/\mu$ as a dimensionless infection rate. A schematic
of our model is shown in Fig.\ref{fig1}.

We adopt a random sequential-update algorithm to simulate the model
\cite{chowdhury2005physics}. We discretize the time in small time
steps $\Delta t$. A node $i$ is first chosen randomly and is tried
to update its state. If node $i$ is susceptible, (s)he becomes
infected with the probability $R_{inf} \Delta t$. If node $i$ is
infected, (s)he recovers to be susceptible with the probability $\mu
\Delta t$. Time is then incremented by $\Delta t/N$ and we iterate
up to some final time. The selection of $\Delta t$ is delicate. Too
small $\Delta t$ will lead to the occurrence of null events very
frequently, so that the simulation becomes inefficient. Too large
$\Delta t$ will cause the updating probabilities larger than one
that are unphysical. In practice, we used $\Delta t=1/(k_{max}
\lambda)$ to minimize the probability that nothing happens while
keeping all probabilities smaller than one, where $k_{max}$ is the
maximal degree of nodes in two layers. Note that the random
sequential-update algorithm has been widely used to simulate the
continuous-time Markov process. It has also been verified that this
algorithm did not produce essential difference from more
sophisticated, but computationally demanding, exact Gillespie
algorithm \cite{gillespie1977exact}.

\section{Simulation Results}

\begin{figure}
\centerline{\includegraphics*[width=1.0\columnwidth]{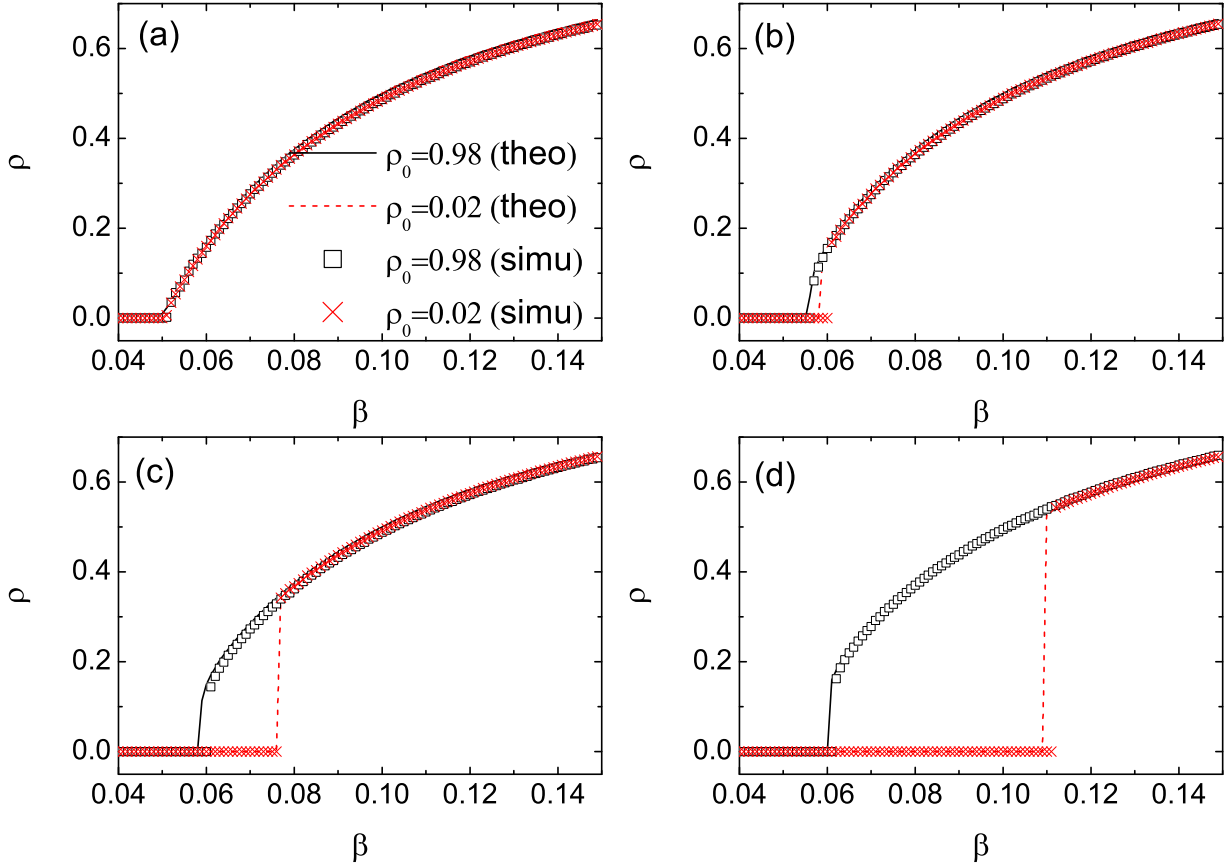}}
\caption{The density of infected nodes $\rho$ as a function of the
infection rate $\beta$ in a two-layer network consisted of two
Poisson random graphs. Two different initial infection densities are
used: $\rho_0=0.98$ (squares) and $\rho_0=0.02$ (crosses). From (a)
to (d) the overlap parameter $\mathcal {O}$ are 1.0, 0.8, 0.5, and
0.2, respectively. The other parameters are $N=10000$ and
$\left\langle k \right\rangle=20$. The lines denote the results of
homogeneous mean-field theory (see Eq.(\ref{eq17})). \label{fig2}}
\end{figure}

We first consider the case where two layer networks are consisted of
Poisson random graphs \cite{ERgraph} with $N=10000$ nodes and the
same average degree $\left\langle k \right\rangle=20$.
Fig.\ref{fig2} shows the simulation results with two different
initial infection density $\rho_0=0.02$ and $\rho_0=0.98$ and
several different values of $\mathcal {O}$, where we have defined
$\rho(t)=N^{ - 1} \sum\nolimits_{i=1}^N {\sigma _i(t)}$. For
$\mathcal {O}=1$, our model recovers to the usual SIS model in
single-layer networks, and the system undergoes a continuous
second-order phase transition from a healthy phase to an endemic
phase as $\beta$ increases, separated by a threshold value of
$\beta_c$ (see Fig.\ref{fig2}(a)). Strikingly, the nature of phase
transition is essentially changed to be discontinuous for $\mathcal
{O}<1$, as shown in Fig.\ref{fig2}(b-d). The results for different
initial conditions do not coincide in a certain range of $\beta\in
\left[\beta_F, \beta_C\right]$, forming a hysteresis region that is
a typical characteristic of a first-order phase transition. Within
the hysteresis region, the system is bistable. Specially, when the
initial density of infection is low, the epidemic will become
extinct. While for high initial density of infection, the system
will maintain a certain proportion of prevalence. As $\mathcal {O}$
decreases, $\beta_F$ is almost unchanged and $\beta_C$ shifts to a
larger value, thus the bistable region is enlarged.

\begin{figure}
\centerline{\includegraphics*[width=1.0\columnwidth]{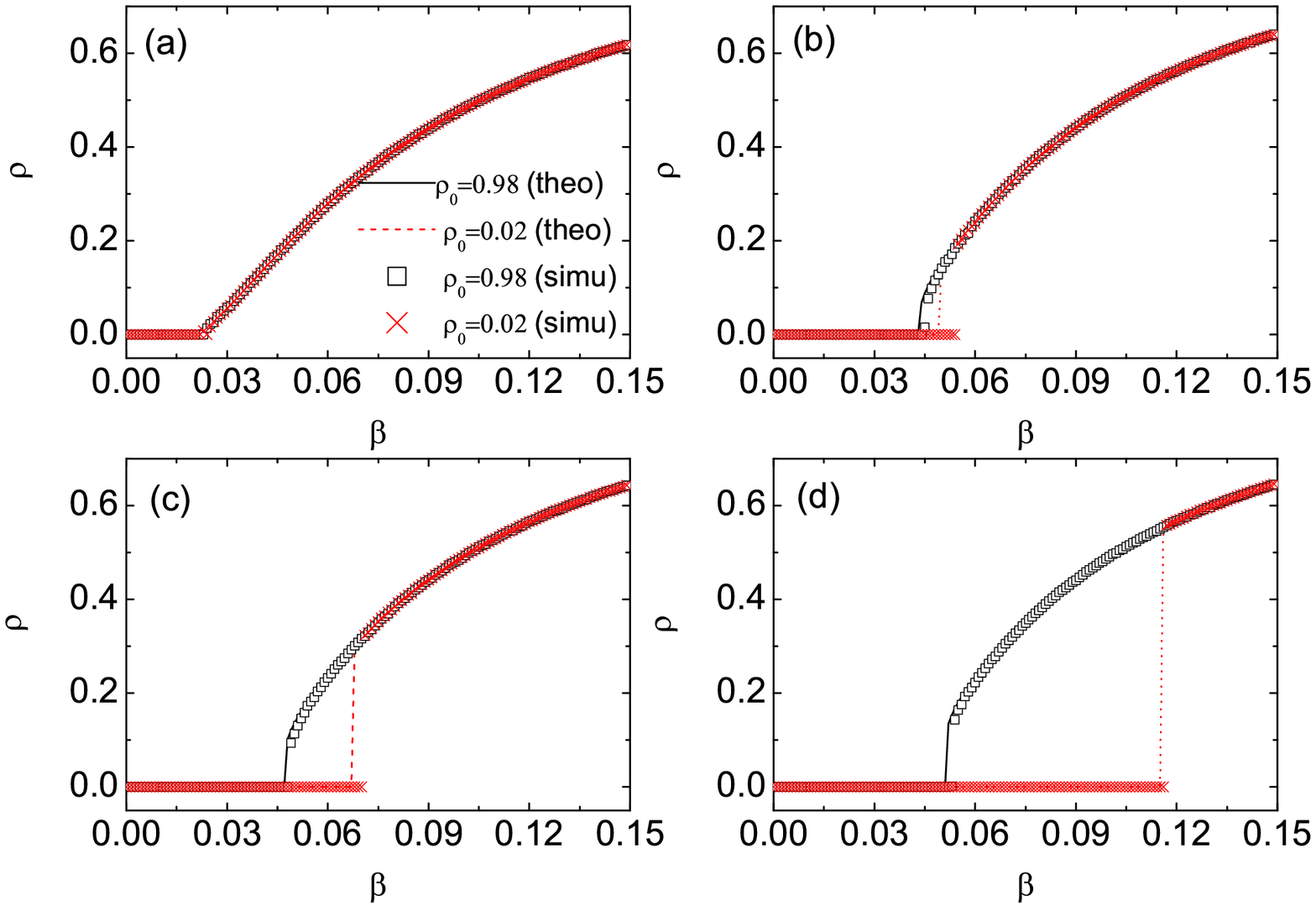}}
\caption{ The density of infected nodes $\rho$ as a function of the
infection rate $\beta$ in a two-layer network, in which the first
layer is a BA network and the second one is obtained by rewiring
edges from a BA network same as the first layer. Two different
initial infection densities are used: $\rho_0=0.98$ (squares) and
$\rho_0=0.02$ (crosses). From (a) to (d) the overlap parameter
$\mathcal {O}$ are 1.0, 0.2, 0.1, and 0.0, respectively. The other
parameters are $N=10000$ and $\left\langle k \right\rangle=20$. The
lines denote the results of individual-based mean-field theory (see
Eq.(\ref{eq13})). \label{fig3}}
\end{figure}

In Fig.\ref{fig3}, we show $\rho$ as a function of $\beta$ in a
two-layer network, in which the first layer is a Barab\'asi-Albert
(BA) network \cite{Science.286.509} and the second one is obtained
by rewiring edges from a BA network same as the first layer. The
qualitative results are the same as Fig.\ref{fig2}. That is to say,
for a more degree-heterogeneous network we also observe the
discontinuous phase transition for the onset of epidemic outbreak
and a bistable region with the coexisting healthy phase and endemic
phase in a more degree-heterogeneous network. However, to observe
such phenomena explicitly, we need to use lower degrees of overlap
in edges among layers.

\begin{figure}
\centerline{\includegraphics*[width=1.0\columnwidth]{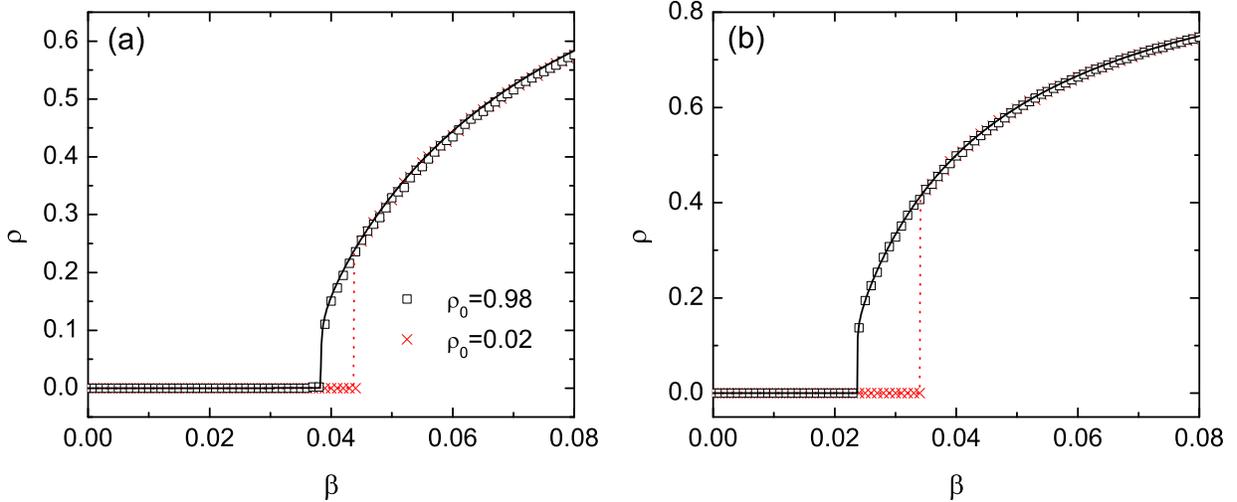}}
\caption{ The density of infected nodes $\rho$ as a function of the
infection rate $\beta$ in a two-layer Poisson random graph, in which
the first layer is completely overlapped with the second one but not
the vice versa. The average degree in the first layer is fixed at
20, and the average degree in the second layer is twice (a) and four
times (b) larger than that of the first layer. Two different initial
infection densities are used: $\rho_0=0.98$ (squares) and
$\rho_0=0.02$ (crosses). The other parameter is $N=10000$. The lines
denote the results of homogeneous mean-field theory (see
Eq.(\ref{eqr17})). \label{fig3_2}}
\end{figure}

\textbf{We now consider the case when the number of edges in the two
layers are not the same. A particular example of interest is that
one layer is completely embedded in another layer. This architecture
will yield one layer completely overlapping with the second one but
not the vice versa. In Fig. \ref{fig3_2}, we show the results in two
Poisson random graphs with $N=10000$ nodes. The average degree in
the first layer is fixed at 20, and the average degree in the second
layer is twice (a) and four times (b) larger than that of the first
layer. One can see that the phase transition is discontinuous. If
the difference of connection densities between the two layers
becomes larger, the discontinuous characteristic of the phase
transition will become more obvious. }

\textbf{The key of the discontinuous phase transitions lies in the
coexistence of two or more different stable phases. The origin of
such a discontinuity in our model stems from the interaction between
the nonlinearity of spreading dynamics introducing by Eq.(\ref{eq2})
and the overlapping among the layers. For a multi-layer network with
low overlap, an intuitive argument with regard to the coexistence of
healthy phase and endemic phase may be presented as follows. For a
high initial density of the infected nodes, most of nodes have at
least one infected neighbor in each layer, such that the spread of
epidemic is equivalent to that in single-layer network. When the
initial density of the infected nodes is low, the reason why
epidemics cannot spread is that most of nodes do not meet the
condition of spreading dynamics in Eq.(\ref{eq2}). That is to say,
under the latter case, nonlinear effect of spreading dynamics does
react and destroy connected infectious clusters, such that the
epidemic dies out. This is akin to explosive synchronization in
multiplex networks \cite{jalan2019explosive,PhysRevE.99.062305},
which demonstrate discontinuous transition in one layer due to
suppression of formation of giant cluster drawn from the second
layer, either due to frequency mismatch in the mirror nodes
\cite{jalan2019explosive} or due to negative intralayer coupling of
the second layer \cite{PhysRevE.99.062305}. In the next section, we
will present a formulistic interpretation to the discontinuous phase
transition based on a mean-field theory.}

\section{Mean-field theory}
\subsection{individual-based mean-field theory}

\begin{figure}
\centerline{\includegraphics*[width=0.6\columnwidth]{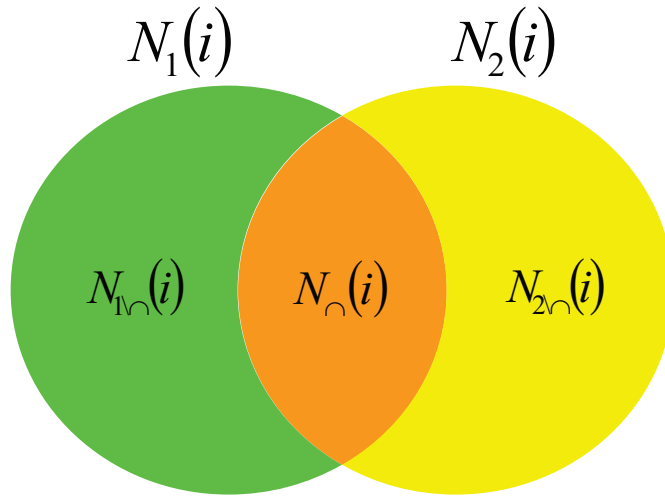}}
\caption{A schematic representation of neighborhood of node $i$.
$N_1(i)$ and $N_2(i)$ denote the sets of neighbors of node $i$ in
the first layer and in the second layer, respectively.
$N_{\cap}(i)=N_1(i) \cap N_2(i)$ is the set of common neighbors of
node $i$ in the two layers. $N_{1(2)\backslash
{\cap}}(i)=N_{1(2)}(i)-N_{\cap}(i)$ is the set of neighbors of node
$i$ belonging to the first (second) layer but not to the second
(first) layer.} \label{fig4}
\end{figure}

To be first, let $\rho_i(t)$ denote the probability of node $i$
being infected at time $t$. That is to say, at time $t$ the state of
node $i$ takes the value $\sigma_i(t)=1$ with the probability
$\rho_i(t)$ and $\sigma_i(t)=0$ with the complementary probability
$1-\rho_i(t)$. To write down the time-evolution equation for node
$i$, a key step is to derive the rate of node $i$ being infected at
time $t$. To do so, we denote by $N_1(i)$ and $N_2(i)$ the set of
neighbors of node $i$ in the first layer and in the second layer,
respectively. Let $N_{\cap}(i)=N_1(i) \cap N_2(i)$ denote the set of
common neighbors of node $i$ in the two layers, such that
$N_{1(2)}(i)=N_{\cap}(i)+N_{1(2)\backslash {\cap}}(i)$, where
$N_{1(2)\backslash {\cap}}(i)$ is the set of neighbors of node $i$
belonging to the first (second) layer but not to the second (first)
layer, see Fig.\ref{fig4} for a schematic. The probability of having
$\{n_1,n_2,n_3\}$ infected neighbors out of the sets $N_{1\backslash
{\cap}}(i)$, $N_{{\cap}}(i)$, and $N_{2\backslash {\cap}}(i)$ can be
repressed as the product of three Poisson binomial distributions,
\begin{eqnarray}
P\left( {{n_1},{n_2},{n_3}} \right) = {P_1}\left( {{n_1}}
\right){P_2}\left( {{n_2}} \right){P_3}\left( {{n_3}}
\right),\label{eq3}
\end{eqnarray}
where
\begin{eqnarray}
{P_1}\left( {{n_1}} \right) = \sum\limits_{Z \in {F_1}}
{\prod\limits_{j \in Z} {{\rho_j}} } \prod\limits_{j \in {Z^c}}
{\left( {1 - {\rho_j}} \right)}.\label{eq4}
\end{eqnarray}
Here $F_1$ are all the subsets of $N_{1\backslash {\cap}}(i)$
containing $n_1$ elements, and $Z^c$ is the complement of $Z$, i.e.,
$Z^c=N_{1\backslash {\cap}}(i) \backslash Z$. Similarly, we can
write down the expressions of $P_2(n_2)$ and $P_3(n_3)$ which are
not shown here to avoid the duplication. According to
Eq.(\ref{eq2}), the rate of node $i$ being infected at time $t$ can
be written as,
\begin{eqnarray}
{R_{\inf }} = \sum\limits_{{n_1} = 0}^{\left| {{N_{1\backslash
{\cap}}}} (i)\right|} {\sum\limits_{{n_2} = 0}^{\left|
{{N_{\cap}}}(i) \right|} {\sum\limits_{{n_3} = 0}^{\left|
{{N_{2\backslash {\cap}}}}(i) \right|} {P\left( {{n_1},{n_2},{n_3}}
\right)\lambda \left( {\frac{{{n_1} + 2{n_2} + {n_3}}}{2}}
\right)\Theta \left( {{n_1} + {n_2} - 1} \right)\Theta \left( {{n_2}
+ {n_3} - 1} \right)} } },\nonumber \\ \label{eq5}
\end{eqnarray}
where ${\left| {{N_{1\backslash {\cap}}}}(i) \right|}$, ${\left|
{{N_{ {\cap}}}}(i) \right|}$, and ${\left| {{N_{2\backslash
{\cap}}}}(i) \right|}$ are the sizes of the sets of ${N_{1\backslash
{\cap}}}(i)$, ${N_{ {\cap}}}(i)$, and ${N_{2\backslash {\cap}}}(i)$,
respectively. To facilitate the calculation of $R_{\inf}$, we
rewrite Eq.(\ref{eq5}) as,
\begin{eqnarray}
{R_{\inf }} =&& \sum\limits_{{n_1} = 0}^{\left| {{N_{1\backslash
{\cap}}}}(i) \right|} {\sum\limits_{{n_2} = 0}^{\left|
{{N_{\cap}}}(i) \right|} {\sum\limits_{{n_3} = 0}^{\left|
{{N_{2\backslash {\cap}}}}(i) \right|} {P\left( {{n_1},{n_2},{n_3}}
\right)\lambda \left( {\frac{{{n_1} + 2{n_2} + {n_3}}}{2}} \right) -
} } } \sum\limits_{{n_1} = 0}^{\left| {{N_{1\backslash {\cap}}}}(i)
\right|} {P\left( {{n_1},0,0} \right)\lambda \frac{{{n_1}}}{2}  }
\nonumber
\\ &&-\sum\limits_{{n_3} = 0}^{\left| {{N_{2\backslash {\cap}}}}(i) \right|}
{P\left( {0,0,{n_3}} \right)\lambda \frac{{{n_3}}}{2}}.\label{eq6}
\end{eqnarray}
The first term on the right hand side of Eq.(\ref{eq6}) can be
computed as,
\begin{eqnarray}
&&\sum\limits_{{n_1} = 0}^{\left| {{N_{1\backslash {\cap}}}}(i)
\right|} {\sum\limits_{{n_2} = 0}^{\left| {{N_{\cap}}}(i) \right|}
{\sum\limits_{{n_3} = 0}^{\left| {{N_{2\backslash {\cap}}}}(i)
\right|} {P\left( {{n_1},{n_2},{n_3}} \right)\lambda \left(
{\frac{{{n_1} + 2{n_2} + {n_3}}}{2}} \right)} } }  \nonumber \\&=&
\frac{{\lambda }}{2}\left[ {\sum\limits_{{n_1} = 0}^{\left|
{{N_{1\backslash {\cap}}}} (i)\right|} {{n_1}{P_1}\left( {{n_1}}
\right) + \sum\limits_{{n_2} = 0}^{\left| {{N_{\cap}}} (i)\right|}
{2{n_2}{P_2}\left( {{n_2}} \right) + \sum\limits_{{n_3} = 0}^{\left|
{{N_{2\backslash {\cap}}}}(i) \right|} {{n_3}{P_3}\left( {{n_3}}
\right)} } } } \right] \nonumber \\&=& \frac{{\lambda }}{2}\left[
{\left\langle {{n_1}} \right\rangle  + 2\left\langle {{n_2}}
\right\rangle  + \left\langle {{n_3}} \right\rangle }
\right],\label{eq7}
\end{eqnarray}
where
\begin{eqnarray}
\begin{gathered}
  \left\langle {{n_1}} \right\rangle  = \sum\limits_{j \in {N_{1\backslash {\cap}}}} {{\rho _j}}  = \sum\limits_j {A_{ij}^1\left( {1 - A_{ij}^2} \right){\rho _j}},  \hfill \\
  \left\langle {{n_2}} \right\rangle  = \sum\limits_{j \in {N_{\cap}}} {{\rho _j}}  = \sum\limits_j {A_{ij}^1A_{ij}^2{\rho _j}},  \hfill \\
  \left\langle {{n_3}} \right\rangle  = \sum\limits_{j \in {N_{2\backslash {\cap}}}} {{\rho _j}}  = \sum\limits_j {A_{ij}^2\left( {1 - A_{ij}^1} \right){\rho _j}}.  \hfill \\
\end{gathered}\label{eq8}
\end{eqnarray}

The second term and the third term on the right hand side of
Eq.(\ref{eq6}) can be computed as,
\begin{eqnarray}
\sum\limits_{{n_1} = 0}^{\left| {{N_{1\backslash {\cap}}}}(i)
\right|} {P\left( {{n_1},0,0} \right)\lambda \frac{{{n_1}}}{2} =
\frac{{\lambda }}{2}} {P_2}\left( 0 \right){P_3}\left( 0
\right)\sum\limits_{{n_1} = 0}^{\left| {{N_{1\backslash {\cap}}}}
(i)\right|} {{n_1}{P_1}\left( {{n_1}} \right) = \frac{{\lambda
}}{2}{P_2}\left( 0 \right){P_3}\left( 0 \right)\left\langle {{n_1}}
\right\rangle },\label{eq9}
\end{eqnarray}
and
\begin{eqnarray}
\sum\limits_{{n_3} = 0}^{\left| {{N_{2\backslash {\cap}}}}
(i)\right|} {P\left( {0,0,{n_3}} \right)\lambda \frac{{{n_3}}}{2}} =
\frac{{\lambda }}{2}{P_1}\left( 0 \right){P_2}\left( 0
\right)\sum\limits_{{n_3} = 0}^{\left| {{N_{2\backslash {\cap}}}}
(i)\right|} {{n_3}{P_3}\left( {{n_3}} \right)}  = \frac{{\lambda
}}{2}{P_1}\left( 0 \right){P_2}\left( 0 \right)\left\langle {{n_3}}
\right\rangle,\label{eq10}
\end{eqnarray}
respectively. Here
\begin{eqnarray}
\begin{gathered}
  {P_1}\left( 0 \right){P_2}\left( 0 \right) = \prod\limits_{j \in {N_1(i)}} {\left( {1 - {\rho _j}} \right)}  = \prod\limits_j {\left( {1 - A_{ij}^1{\rho _j}} \right)}  \hfill, \\
  {P_2}\left( 0 \right){P_3}\left( 0 \right) = \prod\limits_{j \in {N_2(i)}} {\left( {1 - {\rho _j}} \right)}  = \prod\limits_j {\left( {1 - A_{ij}^2{\rho _j}} \right)}  \hfill. \\
\end{gathered}\label{eq11}
\end{eqnarray}
Substituting
Eqs.(\ref{eq7},\ref{eq8},\ref{eq9},\ref{eq10},\ref{eq11}) into
Eq.(\ref{eq6}), we have
\begin{eqnarray}
{R_{\inf }} = && \frac{\lambda }{2}\sum\limits_j {\left( {A_{ij}^1 +
A_{ij}^2} \right)} {\rho _j} - \frac{\lambda }{2}\prod\limits_j
{\left( {1 - A_{ij}^2{\rho _j}} \right)} \sum\limits_j
{A_{ij}^1\left( {1 - A_{ij}^2} \right){\rho _j}} \nonumber  \\ &&-
\frac{\lambda }{2}\prod\limits_j {\left( {1 - A_{ij}^1{\rho _j}}
\right)} \sum\limits_j {A_{ij}^2\left( {1 - A_{ij}^1} \right){\rho
_j}}.\label{eq12}
\end{eqnarray}
Thus, the time-evolution of $\rho_i$ can be written as
\begin{eqnarray}
\begin{gathered}
  \frac{{d{\rho _i}}}{{dt}} =  - \mu {\rho _i} + \left( {1 - {\rho _i}} \right) {R_{\inf
  }}.
\end{gathered}\label{eq13}
\end{eqnarray}

Eq.(\ref{eq13}) is the main theoretical result of the present work.
It is not hard to check that $\rho_i=0$ ($i=1,\cdots,N$) is always a
set of stationary solution of Eq.(\ref{eq13}). Near the onset of
epidemic outbreak, $\rho_i\simeq0$, Eq.(\ref{eq13}) can be
linearized as,
\begin{eqnarray}
\frac{{d{\rho _i}}}{{dt}} =  - \mu {\rho _i} + \lambda \sum\limits_j
{A_{ij}^1A_{ij}^2} {\rho _j},\label{eq14}
\end{eqnarray}
or in the matrix form,
\begin{eqnarray}
\frac{{d\vec \rho }}{{dt}} = \left( { - \mu \textbf{I} + \lambda
 \tilde {\textbf{A}}} \right)\vec \rho,\label{eq15}
\end{eqnarray}
where $\vec \rho  = {\left[ {{\rho _1}, \ldots ,{\rho _N}}
\right]^T}$, $\textbf{I}$ is the $N$-dimensional identity matrix,
and the entries of $\tilde {\textbf{A}}$ are $\tilde A_{ij}=
{A_{ij}^1A_{ij}^2}$. That is to say, $\tilde A_{ij}=1$ only when
$A_{ij}^1=1$ and $A_{ij}^2=1$ simultaneously, and therefore we call
$\tilde {\textbf{A}}$ the overlapping adjacency matrix of multiplex
network. The solution $\rho_i=0$ loses its stability when the
largest eigenvalue of $- \mu \textbf{I} + \lambda \tilde
{\textbf{A}}$ is larger than zero, which determines the epidemic
threshold that is the reciprocal of the largest eigenvalue of $
\tilde {\textbf{A}}$, i.e.,
\begin{eqnarray}
{\beta _C} = \frac{1}{{{\Lambda _{\max }}\left( {\tilde
{\textbf{A}}} \right)}}.\label{eq16}
\end{eqnarray}
For $\mathcal {O}=1$, Eq.(\ref{eq16}) recovers to the result of
single-layer networks \cite{Wang2003,Mieghem2009,Gomez2010},
$\beta_C(\mathcal {O}=1)=1/\Lambda_{max}(\textbf{A})$. For $\mathcal
{O}=0$, $\tilde {\textbf{A}}$ becomes a null matrix and therefore
$\beta_C(\mathcal {O}=0)=\infty$. For $0<\mathcal {O}<1$, $\beta_C$
falls between $\beta_C(\mathcal {O}=1)$ and $\beta_C(\mathcal
{O}=0)$.

\begin{figure}
\centerline{\includegraphics*[width=0.7\columnwidth]{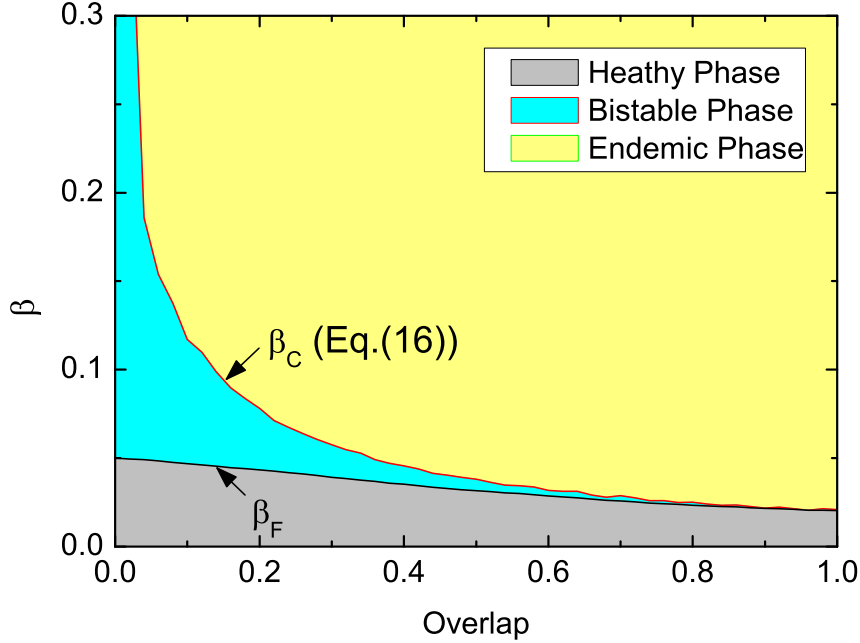}}
\caption{Phase diagram in $\beta \sim \mathcal {O}$ plane. The used
networks are the same as those in Fig.3.  \label{fig5}}
\end{figure}

In Fig.\ref{fig5}, we show the phase diagram of the model in $\beta
\sim \mathcal {O}$ space. We use the same networks as the
Fig.\ref{fig3}. The phase diagram is divided into three regions,
separated by two transition values of $\beta$, $\beta_F$ and
$\beta_C$. $\beta_C$ is obtained by calculating the largest
eigenvalue of the overlapping adjacency matrix (see
Eq.(\ref{eq16})). Note that due to nonlinear characteristic of
Eq.(\ref{eq13}) $\beta_F$ cannot be obtained in general by
analytical derivation. Alternatively, $\beta_F$ is obtained by
numerically solving the steady equation of $\rho_i$ (letting $d
\rho_i/dt=0$ in Eq.(\ref{eq13})) using the initial condition
$\rho_i(0)=1$.

\subsection{homogeneous mean-field theory}
For homogeneous networks, each node is assumed to be statistically
equivalent, and thus $\rho_i=\rho$ for $\forall i$, and degrees of
each node in each layer are the same, i.e., $\sum\nolimits_j
{A_{ij}^1}  = \left\langle k \right\rangle_1$ and $ \sum\nolimits_j
{A_{ij}^2}  = \left\langle k \right\rangle_2$ for $\forall i$. Here,
the rate equation for homogeneous mean-field theory does not need to
be rederived. Alternatively, it can be obtained by rewriting
Eq.(\ref{eq13}) based on the above assumption in homogeneous
networks.\textbf{ Thus, $\sum\nolimits_j {A_{ij}^{1(2)}} {\rho _j}
= \left\langle k \right\rangle_{1(2)} \rho$, $\prod\nolimits_j
\left( 1 - A_{ij}^{1(2)}{\rho_j} \right) = \left( {1 - \rho }
\right)^{\left\langle k \right\rangle_{1(2)} }$, and
$\sum\nolimits_j A_{ij}^{1(2)}\left( {1 - A_{ij}^{2(1)}}
\right){\rho _j} = \left\langle k \right\rangle_{1(2)} \left( {1 -
\mathcal {O}_{1(2)}} \right)\rho$, and Eq.(\ref{eq13}) can be
rewritten as,
\begin{eqnarray}
\frac{{d\rho }}{{dt}} =  - \mu \rho  + \frac{\lambda }{2}\rho \left(
{1 - \rho } \right)\left[ {{{\left\langle k \right\rangle }_1} +
{{\left\langle k \right\rangle }_2} - {{\left\langle k \right\rangle
}_1}{{\left( {1 - \rho } \right)}^{{{\left\langle k \right\rangle
}_2}}}\left( {1 - {\mathcal {O}_1}} \right) - {{\left\langle k
\right\rangle }_2}{{\left( {1 - \rho } \right)}^{{{\left\langle k
\right\rangle }_1}}}\left( {1 - {\mathcal {O}_2}} \right)} \right],
\nonumber
\\\label{eqr17}
\end{eqnarray}
where we have defined $\mathcal {O}_{1(2)}=\sum\nolimits_{i < j}
{A_{ij}^1} A_{ij}^2/\sum\nolimits_{i < j} {A_{ij}^{1(2)}}$ as the
fraction of the number of overlapping edges in the total number of
edges in the first (second) layer. When the numbers of edges in each
layer are the same as considered before, $\mathcal {O}_1=\mathcal
{O}_2=\mathcal {O}$, $\left\langle k \right\rangle_1=\left\langle k
\right\rangle_2=\left\langle k \right\rangle$, Eq.(\ref{eqr17}) can
be simplified to}
\begin{eqnarray}
\frac{{d\rho }}{{dt}} =  - \mu \rho  + \lambda \left\langle k
\right\rangle \left( {1 - \rho } \right)  \rho\left[ {  1- \left( {1
- \mathcal {O}} \right){{\left( {1 - \rho } \right)}^{\left\langle k
\right\rangle }} } \right].\label{eq17}
\end{eqnarray}
Notice that $\rho=0$ is always a stationary solution of
Eq.(\ref{eq17}). Such a trivial solution corresponds to the healthy
phase where no infected nodes survive. According to linear stability
analysis, the solution becomes unstable when the derivative of the
right hand side of Eq.(\ref{eq17}) with respect to $\rho$ at
$\rho=0$ is larger than zero, which determines the epidemic
threshold $\beta_C$,
\begin{eqnarray}
{\beta _C} = \frac{1}{{\left\langle k \right\rangle \mathcal
{O}}}.\label{eq18}
\end{eqnarray}

\begin{figure}
\centerline{\includegraphics*[width=1.0\columnwidth]{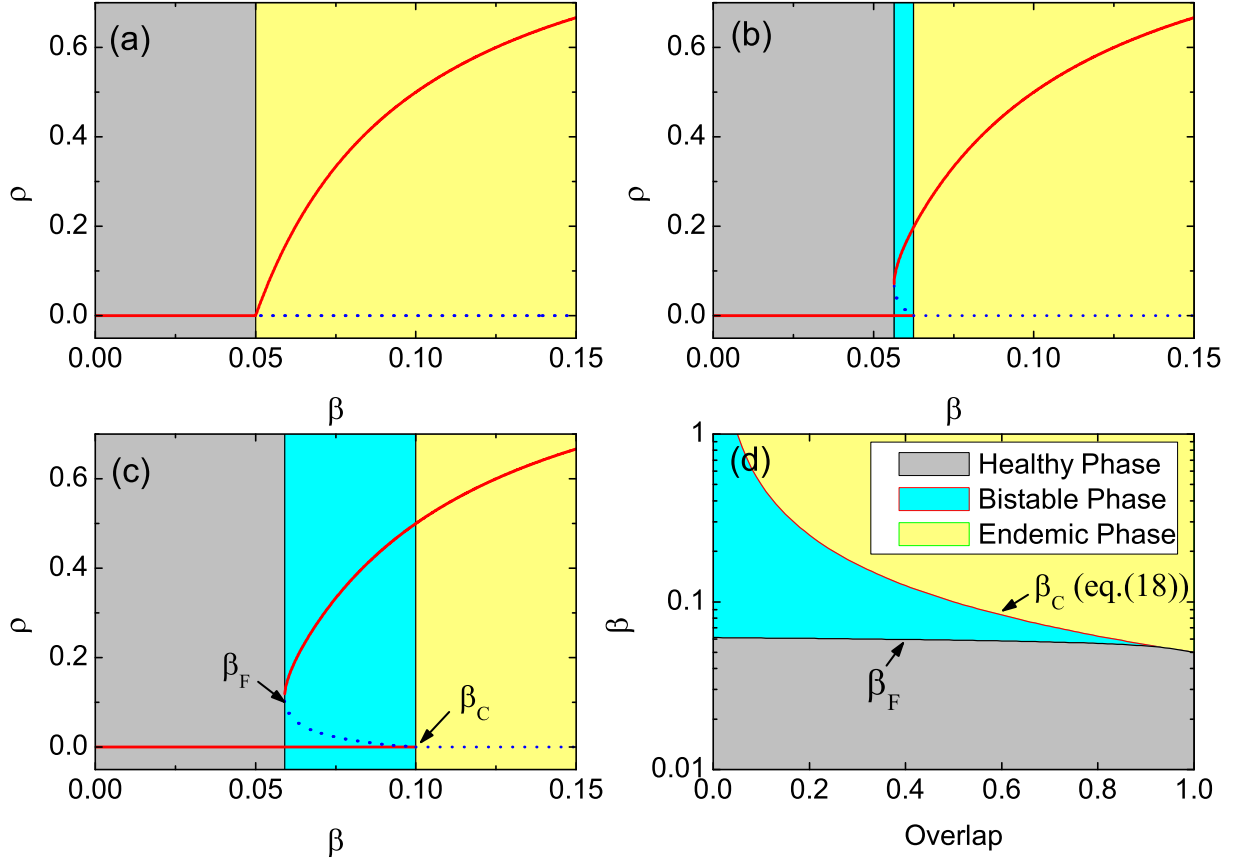}}
\caption{Results from the homogeneous mean-field theory. (a-c) shows
$\rho$ versus $\beta$ for three different $\mathcal {O}$: 1.0 (a),
0.8(b), and 0.5 (c). (d) shows the phase diagram in $\beta \sim
\mathcal {O}$ plane. Solid lines and dotted lines in (a-c) denote
stable and unstable solutions, respectively. The average degree is
$\left\langle k \right\rangle=20$.  \label{fig6}}
\end{figure}

Comparing to mean-field equation of the SIS model in single-layer
networks \cite{RevModPhys.87.925}, our model can give rise to an
additional term in Eq.(\ref{eq17}), ${  \left( {1 - \mathcal {O}}
\right){{\left( {1 - \rho } \right)}^{\left\langle k \right\rangle
}} }$. Obviously, the additional term vanishes in the case of
$\mathcal {O}= 1$. Importantly, we shall see that for $\mathcal
{O}\neq 1$ the additional term can lead to an essential change in
the bifurcation of the model. The results for the simple mean-field
theory are summarized in Fig.\ref{fig6}. Fig.\ref{fig6}(a-c) shows
$\rho$ as a function of $\beta$ for three distinct values of
$\mathcal {O}$. For $\mathcal {O}=1$, our model recovers to the
standard SIS model, and it is well-known that $\rho$ shows a
transcritical bifurcation as $\beta$ varies. Across the epidemic
threshold $\beta=1/\left\langle k \right\rangle$ (here we have used
$\left\langle k \right\rangle=20$) from below, the trivial solution
$\rho=0$ loses its stability, and a new solution of $\rho \neq 0$
arises. In physics, we call that the model undergoes a continuous
phase transition from a healthy phase ($\rho=0$) to an endemic phase
($\rho>0$) at $\beta=\beta_C$. For $\mathcal {O}\neq 1$, the
bifurcation feature is changed essentially. When $\beta<\beta_F$,
$\rho=0$ is only stable solution. When $\beta_F<\beta<\beta_C$,
there exist two stable solutions, $\rho=0$ and $\rho>0$, and an
unstable solution ($\rho^{uns}$) lying in between the two stable
solutions. Depending on the initial density $\rho_0$ of infected
nodes, the system will arrive at either a healthy phase (for
$\rho_0<\rho^{uns}$) or an endemic phase (for $\rho_0>\rho^{uns}$).
As $\beta$ approaches $\beta_F$ or $\beta_C$, one of stable
solutions and the unstable solution of $\rho$ get close to each
other, until they colloid and annihilate via a saddle-node
bifurcation. When $\beta>\beta_C$, $\rho=0$ is unstable and $\rho>0$
is only stable. Therefore, for $\mathcal {O}<1$ the system is
divided into three phases in terms of $\beta$. For $\beta<\beta_F$
the system is in the healthy phase. For $\beta>\beta_C$ the system
is in the endemic phase. Between them, the system is in bistable
phase. Fig.\ref{fig6}(d) shows the phase diagram in the parametric
space $\beta \sim \mathcal {O}$. The boundary line $\beta_F$ shows a
very slow decrease as $\mathcal {O}$ increase, and the other one
$\beta_C$ decreases obviously with $\mathcal {O}$ according to
Eq.(\ref{eq18}), such that we can see that the bistable region is
clearly enlarged as $\mathcal {O}$ decreases.

\section{Comparison between simulation and theory}
It is expected that the homogenous mean-field theory coincides with
the simulation results in Poisson random graph (shown in
Fig.\ref{fig2}). To compare them, we numerically solve
Eq.(\ref{eq17}) using the same initial conditions as the
simulations, and theoretical results are shown by lines in Fig.2.
There are an excellent agreement between the theory and simulation.
We should note that the theoretical value of $\beta_C$ is not easy
to access in simulation. For example, for $\left\langle k
\right\rangle=20$ and $\mathcal {O}=0.5$, we have $\beta_C=0.1$ in
terms of Eq.(\ref{eq18}) (shown in Fig.\ref{fig6}(c)). In
simulation, we use $\rho_0=0.02$ and give $\beta_C=0.076$, as shown
in Fig.\ref{fig2}(c). In principle, we can access the theoretical
limit of Eq.(\ref{eq18}) by using a lower initial density of
infection in simulation. However, if the number of infected seeds is
very small, the finite-size fluctuations may drive, with a very high
probability, the system to the absorbing state whenever no more
infected nodes survive. Once the absorbing state is reached, the
system cannot be left. Therefore, in order to verify the theoretical
prediction in Eq.(\ref{eq18}) with an adequate accuracy, one needs
to use a considerable large network size to reduce the finite-size
fluctuations. It will certainly increase more computational
resource.

For more degree heterogeneous networks, individual-based mean-field
theory is more appropriate. Using the duplex networks same as those
in Fig.\ref{fig3}, we numerically solve Eq.(\ref{eq13}) to obtain
stationary value of $\rho_i$ and the average infection density
$\rho=N^{-1}\sum\nolimits_{i=1}^N {\rho _i}$, as indicated by lines
in Fig.\ref{fig3}. As expected, the theory can well reproduce the
simulation results.

\section{Results on other multiplex networks}

\begin{figure}
\centerline{\includegraphics*[width=1.0\columnwidth]{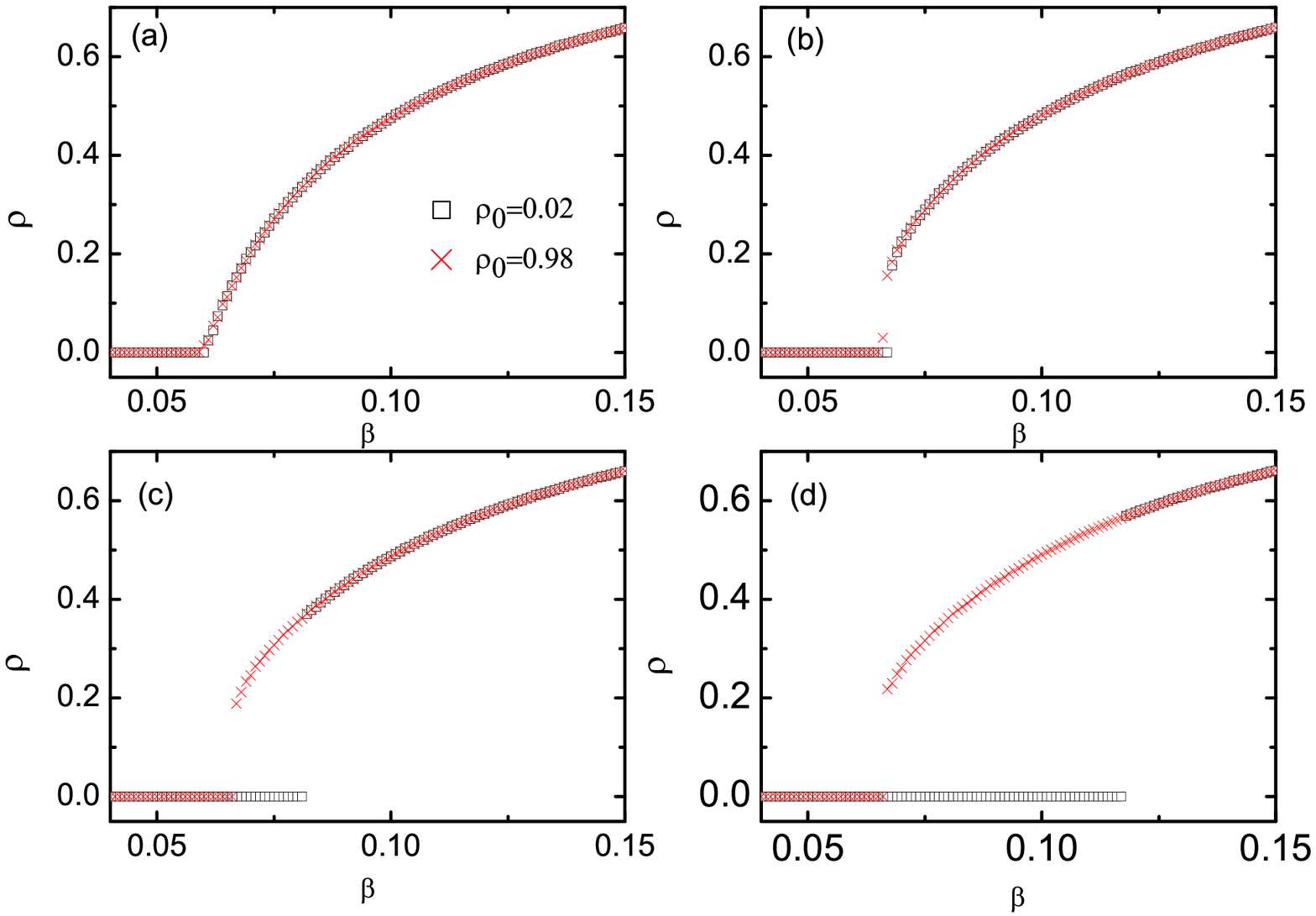}}
\caption{The density of infected nodes $\rho$ as a function of the
infection rate $\beta$ in a two-layer network where the first layer
is consisted of a Watts-Strogatz small-world network and the second
layer is obtained by randomly rewiring the first layer network such
that a given fraction of overlapping edges is achieved. Two
different initial infected densities are used: $\rho_0=0.02$
(squares) and $\rho_0=0.98$ (crosses). From (a) to (d) the overlap
parameters $\mathcal {O}$ are 1.0, 0.8, 0.5, and 0.2, respectively.
The other parameters are $N = 10000$ and $\left\langle k
\right\rangle=20$. \label{fig7}}
\end{figure}

\begin{figure}
\centerline{\includegraphics*[width=1.0\columnwidth]{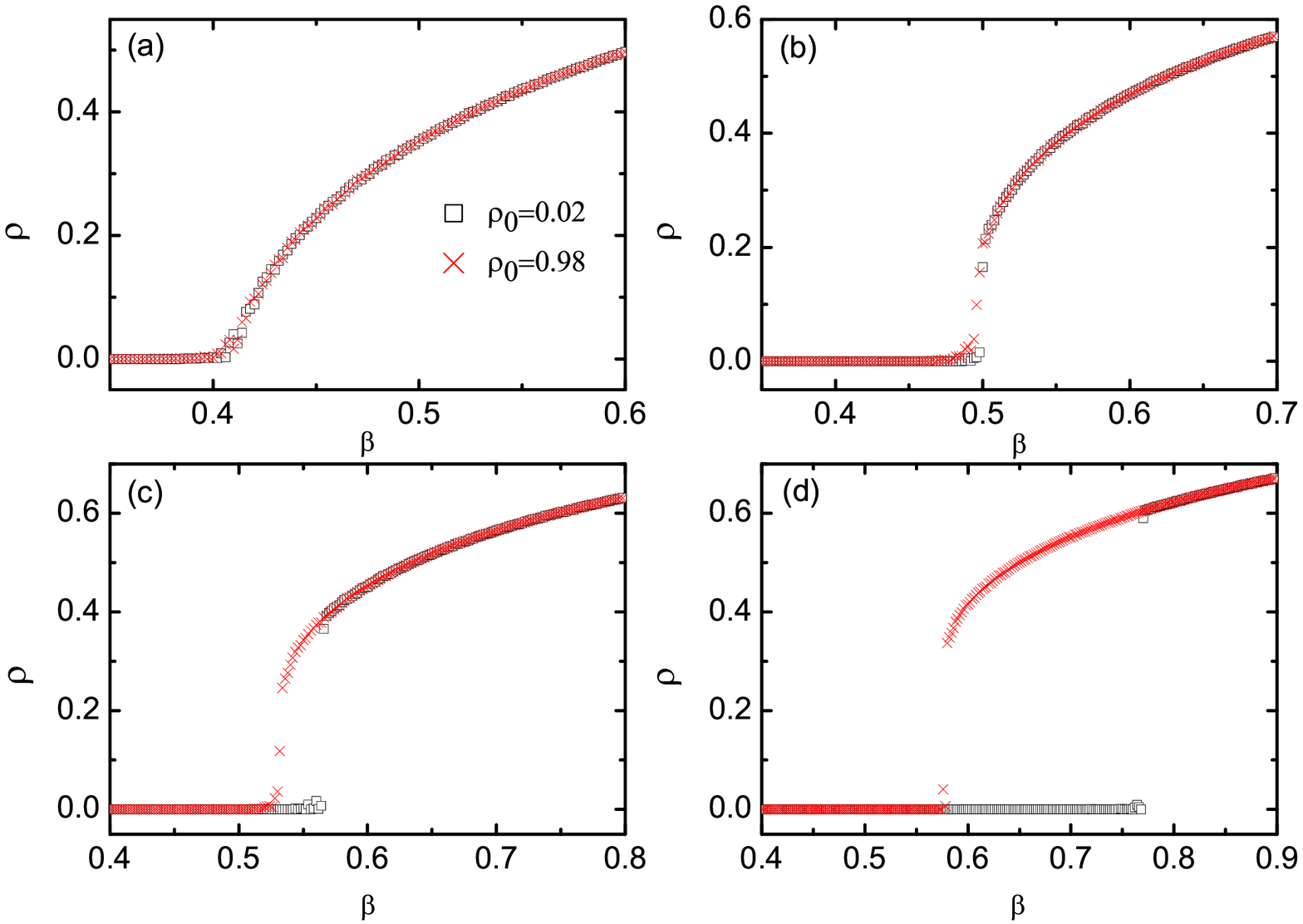}}
\caption{The density of infected nodes $\rho$ as a function of the
infection rate $\beta$ in a two-layer network where the first layer
is consisted of a $100\times100$ square lattice and the second layer
is obtained by randomly rewiring the first layer network such that a
given fraction of overlapping edges is achieved. Two different
initial infected densities are used: $\rho_0=0.02$ (squares) and
$\rho_0=0.98$ (crosses). From (a) to (d) the overlap parameters
$\mathcal {O}$ are 1.0, 0.8, 0.7, and 0.5, respectively.
\label{fig8}}
\end{figure}

To validate the generality of our conclusion, we also present the
simulation results in other multiplex networks. In Fig.7, the first
layer is consisted a Watts-Strogatz small-world network
\cite{watts1998collective}. The small-world network is generated as
follows. We start with a regular ring network with $N=10000$ nodes
in which each node is connected to its first $K=20$ neighbors ($K/2$
on either side), and we then randomly rewire each edge of the ring
network with probability $p=0.05$ such that $pNK/2$ long-range edges
are generated. In Fig.8, the first layer is consisted of a
$100\times 100$ square lattice (periodic boundary) in which each
node is connected to its four nearest neighbors. The second layers
both in Fig.7 and Fig.8 are obtained by randomly rewiring the first
layer network such that a given fraction of overlapping edges is
achieved. From Fig.7 and Fig.8, one sees that for less degrees of
overlapping edges in the two layers, a discontinuous phase
transition can be also observed. That is to say, the main conclusion
in our work holds for other network models as well.

In Fig.9, we show the results on a three-layer network consisted of
three Poisson random graphs with $N=10000$ and $\left\langle k
\right\rangle=20$, in which we have assumed that a susceptible node
can be infected only when it has at least one infectious neighbor in
each layer. It can be seen that the main conclusions are consistent
with those in a two-layer network. At last, we perform simulations
on two real multiplex networks: C.Elegans multiplex connectome
\cite{chen2006wiring,de2015muxviz} and SACCHCERE multiplex network
\cite{stark2006biogrid,de2015structural}. We find that they can
produce discontinuous phase transition as well, as shown in Fig.10.

\begin{figure}
\centerline{\includegraphics*[width=1.0\columnwidth]{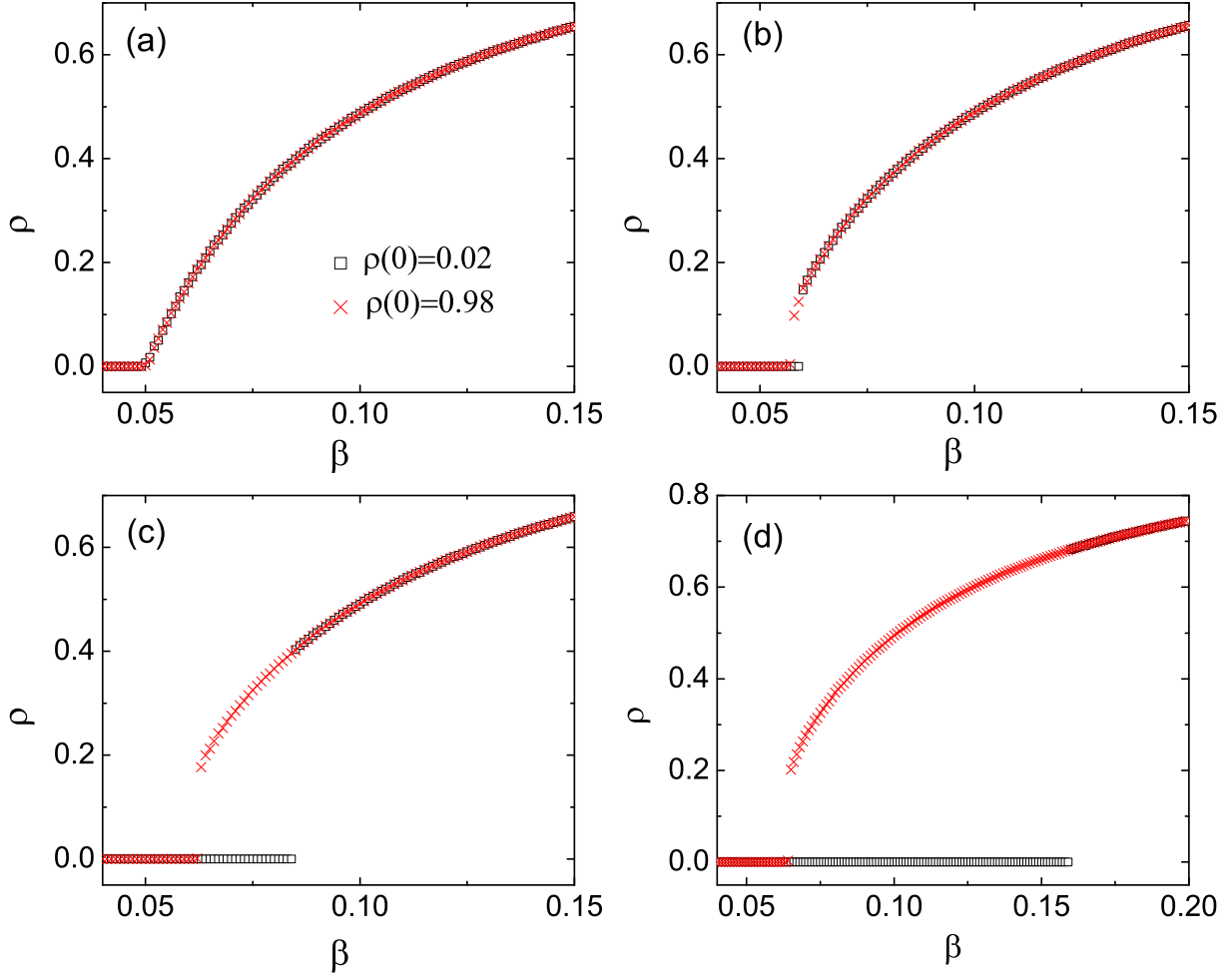}}
\caption{The density of infected nodes $\rho$ as a function of the
infection rate $\beta$ in a three-layer network consisted of three
Poisson random graphs. Two different initial infection densities are
used: $\rho_0=0.02$ (squares) and $\rho_0=0.98$ (crosses). From (a)
to (d) the overlap parameter $\mathcal {O}$ are 1.0, 0.8, 0.5, and
0.2, respectively. The other parameters are $N=10000$ and
$\left\langle k \right\rangle=20$. \label{fig9}}
\end{figure}

\begin{figure}
\centerline{\includegraphics*[width=1.0\columnwidth]{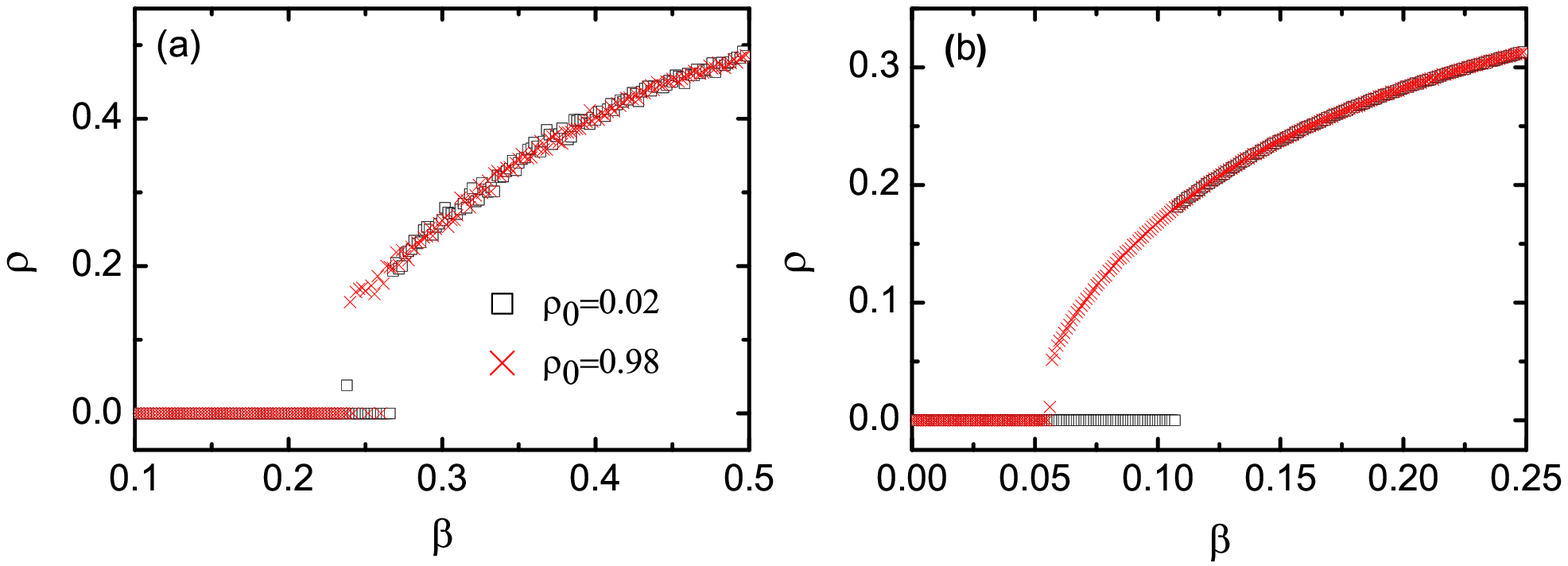}}
\caption{The density of infected nodes $\rho$ as a function of the
infection rate $\beta$ in two real multiplex networks. Two different
initial infection densities are used: $\rho_0=0.02$ (squares) and
$\rho_0=0.98$ (crosses). (a) C.Elegans multiplex connectome
consisted of three layers corresponding to different synaptic
junctions: electric, chemical monadic, and chemical polyadic
\cite{chen2006wiring,de2015muxviz}. The multiplex network contains
279 nodes and 5863 edges . The overlap parameter is $\mathcal
{O}=0.165$. (b) SACCHCERE multiplex network
\cite{stark2006biogrid,de2015structural}. The original network is
consisted of seven layers, but we only used the first four layers:
physical association, suppressive genetic interaction defined by
inequality, direct interaction, and synthetic genetic interaction
defined by inequality. The multiplex network contains 6570 nodes and
151991 edges. The overlap parameter is $\mathcal {O}=0.013$. Note
here that the number of edges in each layer is not the same, and the
overlap parameter $\mathcal {O}$ is defined as the number of
overlapping edges in all layers divided by the minimum of the number
of edges in all layers. \label{fig10}}
\end{figure}

\section{Conclusions}
In conclusion, we have studied an SIS-type epidemic spreading model
in multiplex networks, in which a susceptible individual can be
infected only when (s)he has at least one infectious neighbor in
each layer. We find that the proportion of overlapping edges between
different layers has a significant impact on the nature of phase
transition for the epidemic outbreak. When all the edges are
completely overlapped, the model recovers to the standard SIS model
in single-layer networks, and it undergoes a continuous phase
transition. Otherwise, the model shows an essentially different
nature of phase transition, that is of a discontinuous first order.
Using low and high initial densities of infected individuals, the
model shows two distinct transition pathways from an endemic
extinction phase to an endemic spread phase as the infection rate
increases. Such two pathways form a hysteresis region in which the
system is bistable with the coexisting endemic extinction phase and
endemic spread phase. As the degree of overlapping edges decreases,
the left boundary of the hysteresis region changes slowly, but the
right boundary of the hysteresis region moves swiftly to a larger
value of the infection rate, such that the hysteresis region is
enlarged as $\mathcal {O}$ decreases. Moreover, we have developed an
individual-based mean-field theory that can derive the
time-evolution equations of infected probabilities of individuals.
The individual-based mean-field equations can be reduced to a single
equation of average infection density. Such a coarse graining is
advantageous to unveil the physical mechanics of phenomena observed
in simulations. By linear stability analysis, we have derived the
threshold of epidemic outbreak, corresponding to the right boundary
of hysteresis region. Our theory can well reproduce the simulation
results.

Recently, there were some studies that reported distinct mechanisms
leading to discontinuous or explosive spreading outbreak in
single-layer networks, such as reinfections in social contagions
\cite{gomez2015abrupt}, synergistic effect in transmission rate
\cite{gomez2016explosive}, cooperative coinfections of multiple
diseases \cite{EPL2013,NatPhys2015,PNAS2015}, core contact process
\cite{NJP17.023039,PhysRevE.87.062819}, and higher-order
interactions between individuals \cite{iacopini2019simplicial}, etc.
The present work shows a new mechanism that can lead to a
discontinuous phase transition due to the interacting spreading
dynamics across different network layers. This mechanism underlies
the importance of correlations in edges belong to different layers.
Therefore, our study adds to the continuing effort of the effects of
multiplexity on dynamic processes on multiplex networks, compared to
conventional single-layer ones. On the one hand, in most social
systems, individuals interact with each other in complicated
patterns that include multiple types of relationships. The present
findings may improve our understanding for some real-world spreading
processes in such complex systems such as the spread of a rumor, the
formation of a new opinion. On the other hand, a common
characteristic of discontinuous epidemic outbreak is that
infinitesimal increase of the external parameters, such as infection
rate, can give rise to a considerable macroscopic spreading scope.
There is no doubt that it brings more challenges for controlling or
predicting epidemic outbreaks \cite{d2019explosive}. Finally, we
expect that the present theoretical findings can be supported by
empirical or experimental research in the future.

\begin{acknowledgments}
We acknowledge supports from the National Natural Science Foundation
of China (Grant No. 11875069 and No. 61973001).
\end{acknowledgments}

%

\end{document}